\documentclass[]{spie}  

\usepackage{amsmath,amsfonts,amssymb}

\usepackage{graphicx}
\usepackage[colorlinks=true, allcolors=blue]{hyperref}
\usepackage{booktabs}
\usepackage{array}
\usepackage[table]{xcolor}
\usepackage{float}
\usepackage[utf8]{inputenc} 
\usepackage[T2A]{fontenc} 
\usepackage{orcidlink}

\title{Denoising OCT Images Using Steered Mixture of Experts with Multi-Model Inference}
\author[a,b]{Ayta\c{c} \"Ozkan\,\orcidlink{0000-0002-8188-7824}}
\author[b]{Elena Stoykova}
\author[a]{Thomas Sikora}
\author[b]{Violeta Madjarova}

\affil[a]{Communication Systems Group, Technical University of Berlin, Germany}
\affil[b]{Institute of Optical Materials and Technologies, Bulgarian Academy of Science, Sofia, Bulgaria}

\authorinfo{Further author information: (Send correspondence to A.\"Ozkan)\\A.\"Ozkan: E-mail: aytac.oezkan@tu-berlin.de}

\begin{document} 
\maketitle
\begin{abstract}

In Optical Coherence Tomography (OCT), speckle noise significantly hampers image quality, affecting diagnostic accuracy. Current methods, including traditional filtering and deep learning techniques, have limitations in noise reduction and detail preservation. Addressing these challenges, this study introduces a novel denoising algorithm, Block-Matching Steered-Mixture of Experts with Multi-Model Inference and Autoencoder (BM-SMoE-AE). This method combines block-matched implementation of the SMoE algorithm with an enhanced autoencoder architecture, offering efficient speckle noise reduction while retaining critical image details. Our method stands out by providing improved edge definition and reduced processing time. Comparative analysis with existing denoising techniques demonstrates the superior performance of BM-SMoE-AE in maintaining image integrity and enhancing OCT image usability for medical diagnostics.
\end{abstract}

\keywords{Image denoising, multi-model inference, regression model, optical coherence tomography}

\section{INTRODUCTION}
\label{sec:intro}  
High image quality is an essential requirement in OCT for obtaining accurate and detailed medical images for precise diagnostics. Usage of low coherence light evokes formation of speckle noise due to interference of the mutually coherent light waves reflected from the structures within the imaged sample. When these waves combine, they form random patterns leading to a grainy appearance in the OCT images. Speckle noise can obscure fine details, reduce image quality, and hinder interpretation and analysis of the acquired data. In spite of the advancements in coherent imaging, the problem of the speckle noise persists, significantly impacting quality of medical images and OCT images in particular. Reduction of the speckle noise as a signal dependent noise demands development of sophisticated denoising techniques. 

In general, denoising techniques can be classified into traditional and machine learning based methods. Traditional speckle noise reduction methods such as the non-local mean (NLM) \cite{RN101} and block-matching 3D filtering (BM3D) \cite{RN73} and BMxD \cite{RN78} have demonstrated commendable performance both qualitatively and perceptually. Nevertheless, in NLM and BM3D methods, there is a possibility of losing edge information when the local regions cannot be matched accurately. 

Thanks to the significant progress made in the last decade in deep learning, Convolutional Neural Networks (CNNs) and transformer-based architectures are becoming the state-of-the-art for most image restoration tasks, which leads to their widespread use in medical imaging. Some of the well-known denoising algorithms such as FFDNet,\cite{RN75} DnCNN,\cite{RN76} Noise2Void (N2V)\cite{RN77} and recently published Speckle Split Noise2Void (SSN2V)\cite{RN72}  represent techniques primarily focused on reducing additive noise which is different from the speckle noise typically found in OCT images. An important drawback of algorithms based on neural networks is their heavy reliance on data, particularly on noise-free ground-truth images for training. 

Thus, Traditional and deep learning methods have not fully resolved the speckle noise issue in OCT. In response to this challenge, our paper introduces the BM-SMoE-AE method, which we propose as a new approach for denoising OCT images. This method is expected to preserve sharp edges and smooth transitions. More specifically, we adapt our published denoising framework \cite{RN67} for speckle noise reduction in OCT and medical images by replacing the gradient descent (GD) optimization \cite{RN69} in  the BM-SMoE framework. A revised autoencoder (AE) neural network architecture, as referenced in \cite{RN68}, is employed for the estimation of Gaussian kernel parameters, incorporating a composite loss function to enhance the accuracy of this process. Our implementation of block-matching (BM) in the SMoE framework is designed to accelerate runtime performance and ensure a predictable number of kernel usages, particularly in multi-model inference scenarios. We detail the BM-SMoE-AE method  focusing on edge preservation, block-matching, and autoencoder functionality, and compare our results with the results obtained with other state-of-the-art methods.

The objectives of the work are the following:
\begin{enumerate}
    \renewcommand{\labelenumi}{\roman{enumi}.}
    \item to extend the BM-SMoE model \cite{RN67} for the case of the OCT images by incorporating an adaptive threshold technique.
    \item to adapt the Patch-Based Manifold Learning (PBML)\cite{RN100}-SMoE model for OCT image denoising and to demonstrate its expanded utility and robustness.
    \item to  integrate a composite loss function and an adaptive learning rate scheduler, yielding improved reconstruction quality in PSNR and SSIM metrics over the prior work \cite{RN68} on the BM3D’s  dataset.
    \item to demonstrate superior denoising performance compared to other state of the art methods.
\end{enumerate}
To prove that the proposed novel framework with multi-model inference adequately addresses the problem of OCT image denoising and exhibits significant denoising capacity at enhanced effectiveness in reducing speckle noise compared to the state-of-the-art methods, we used global image quality metrics (PSNR, SSIM, etc.). 

\section{METHODS}
\label{sec:method}
\subsection{The Edge-Aware SMoE Model}

\begin{figure}[H]
   \begin{center}
   \begin{tabular}{c} 
   \includegraphics[height=10cm]{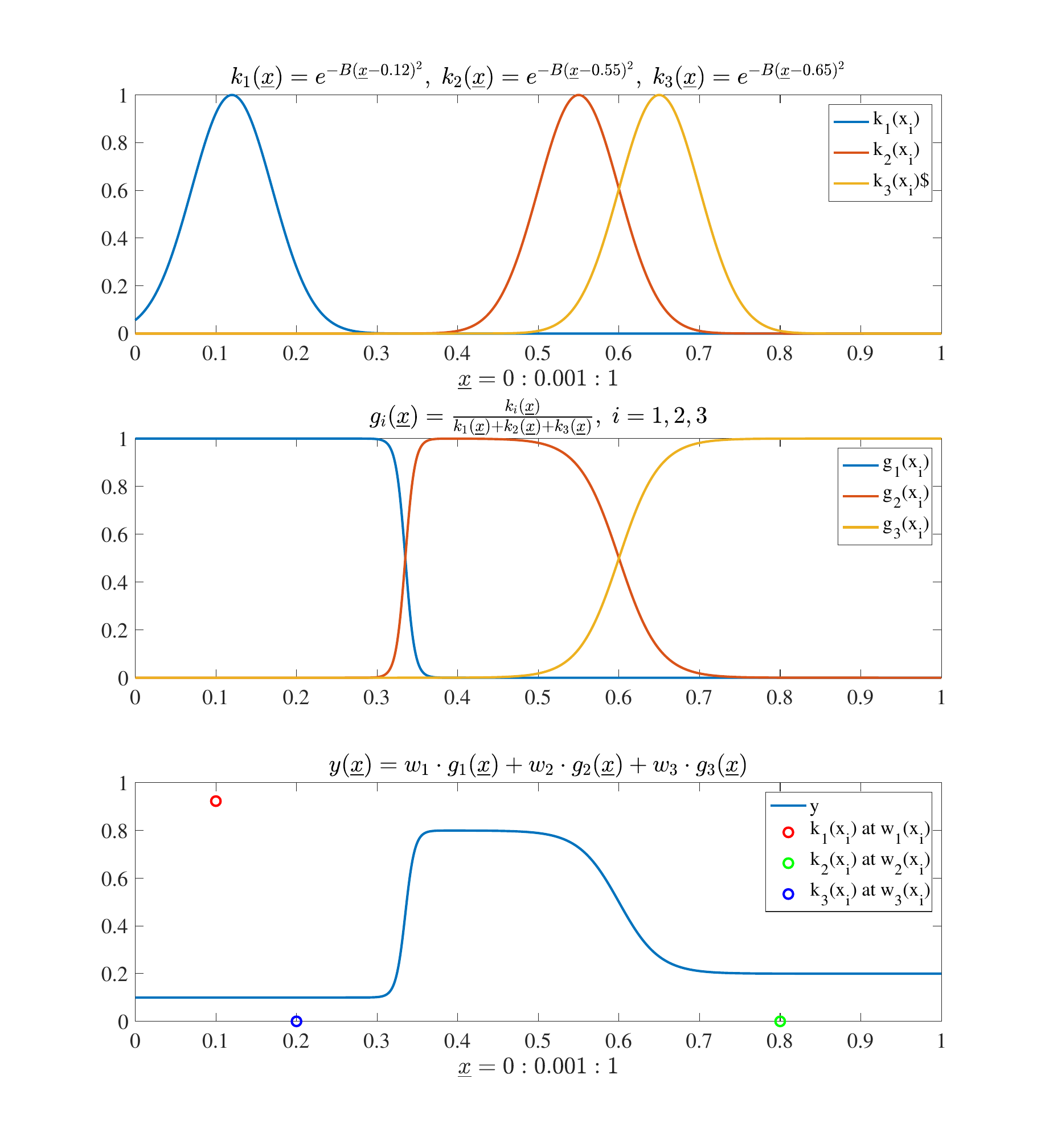}
   \end{tabular}
   \end{center}
   \caption[SMoE] 
   { \label{fig:1D_SMoE} SMoE’s kernels, gates, and regression function respectively.}
   \end{figure}

Before delving into the SMoE model in higher dimensions, we aim to elucidate the fundamental principles underlying the regression framework. This  is demonstrated through a 1D artificial signal for number of three kernels ($N=3$), as depicted in Fig.~\ref{fig:1D_SMoE}.

The SMoE framework’s key component, Gaussian Kernels are fundamental part of our gated regression framework and, as depicted in Fig.~\ref{fig:1D_SMoE}, they can be described as:
\begin{gather}
\label{eq:eq1}
    \underline{x} = 0:0001:1 \nonumber \\
    k_{i} = \exp\left( -\Sigma \cdot (\underline{x} -\mu_{i})^{2} \right) \quad \forall i = 1,2,3
\end{gather}
Here $\underline{x}$ is the sample positions ranging from 0 to 1 in steps of $1\times{10}^{-4}$. The kernel function $k_i$ and the kernel center $\mu_i$ penalize distances away from the local position $\mu_i$, where the approximation is centered. They employ the smoothing parameter $\Sigma$ (the inverse of the kernel bandwidth) to control the strength of this penalty \cite{RN21}. A larger bandwidth results in a smoother, wider kernel that encompasses more data points, while a smaller bandwidth results in a narrower kernel that focuses more on data points close to the center.

The SMoE model incorporates a sparse gating network, essential for determining each kernels' contribution at every data point. This network, depicted in the second row of Fig.~\ref{fig:1D_SMoE}, consists of gating functions $g_1$, $g_2$, and $g_3$ in the considered example. The gating functions are closely linked to their corresponding kernel functions $k_1$, $k_2$, and $k_3$. The network's key feature is its sparsity, achieved by using a minimal set of parameters, specifically the weights $w_1$, $w_2$, and $w_3$, and the means $\mu_1=0.12$, $\mu_2=0.55$, and $\mu_3=0.65$. This design significantly reduces the number of parameters required, enhancing the model's efficiency without compromising its effectiveness.

The sparse and selective activation of these kernels by the gating functions enables the SMoE model to efficiently reconstruct image features with sharp edges and smooth transitions. Such capability is particularly valuable in image processing, where balancing the preservation of fine details with smooth gradient transitions is crucial for high-quality image representation. The 'soft' gating function, a fundamental aspect of the SMoE model, is explained in equation \ref{eq:eq2}.

\begin{equation}
\label{eq:eq2}
\displaystyle g_{i}(\underline{x}) = \frac{k_{i}(\underline{x})}{\sum _{i=1}^{N} k_{i}(\underline{x})}
\end{equation}

This function in Equation \ref{eq:eq2} ensures normalization of the kernels to achieve a partition of unity, where the sum of all kernel contributions at any specific data point equals one. This soft partitioning approach allows for overlapping regions of influence from multiple kernels, enhancing the model's flexibility and representational capacity. Note that to preserve simplicity, not all properties of the SMoE model are presented here; however, they are discussed in detail in Section ~\ref{sec:ae-smoe}.

The last row of Fig.~\ref{fig:1D_SMoE} depicts the regression function that combines the normalized (soft-gated) kernels linearly for the construction of more complex data structures.
\begin{equation}
\label{eq:eq3}
y\left(\underline{x}\right)=\ \sum_{i=1}^{N}{w_i\cdot g_i(\underline{x})}	
\end{equation}
In Equation \ref{eq:eq3}, the weight $w_i$ adjusts how much each gate influences the final output. By changing parameters like $w_i$ and $\mu_i$, we can model different shapes and behaviors. This flexibility is common in kernel methods, such as in Support Vector Machines (SVMs) and Gaussian processes.

The SMoE model differs from these traditional methods because of its unique gating feature. This feature is useful in situations where we need to be careful with computational resources. The SMoE model uses a few parameters to effectively manage complex tasks in image processing, like denoising, compressing, and enhancing image resolution. Its ability to perform these tasks efficiently makes it a useful tool in this field.

\subsection{The Block-Matching Algorithm}
\label{sec:bm-alg}
The BM algorithm has previously been used within various classical\cite{RN73,RN78} and neo-classical denoising techniques\cite{RN74}. Main motivation of its usage is that it transforms real-world noise distributions into pseudo-Gaussian noise, which makes noise more predictable and manageable.\cite{RN89} Also, utilizing a manifold of image patches enables us to employ fewer Gaussian kernels, contributing to enhanced computational efficiency and more precise data representation. 

We implement the block matching to reclaim the certain number of similar patches in OCT image $\ \in\ \mathbb{R}^{height\times width}$. A reference image patch $\mathcal{P}$ with size of $k \times k$ pixels, for example $8 \times 8$ pixels,  in the image according to a given stride step size is selected. To acquire the similar patches, we used the normalized Euclidean quadratic distance between patches, $\displaystyle d(\mathcal{P} ,Q) =\ ||\gamma (\mathcal{P}) - \gamma (Q) ||_{2}^{2}$ where $\gamma$ is a hard thresholding operator containing a threshold $\lambda_{2D}$ and the noise variance $\sigma$.  For $\sigma\le\ 40$, one has $\lambda_{2D} =0.$ We used $\tau^{hard}$ \cite{RN73} for the distance threshold for $d$ under which two patches are assumed similar. $\mathbb{P}\left(P\right)$, denoting 3D group, is the built by staking up the matched patches  $\mathcal{P}$.  Only $N^{hard}$ patches in $\mathcal{P}$ that are closest to the reference patch are kept in the 3D group. These patches are sorted according to their distance to $\mathbb{P}$.
\subsection{The Autoencoder and SMoE Gating Network}
\label{sec:ae-smoe}
Recent research \cite{RN68}, demonstrated that shallow neural networks are more effective at determining kernel parameters compared to GD optimizers\cite{RN69}. Consequently, we've integrated deep neural networks into our multi-model inference algorithms, enhancing their performance and accuracy.

\begin{figure} [ht]
   \begin{center}
   \begin{tabular}{c} 
   \includegraphics[height=3.5cm]{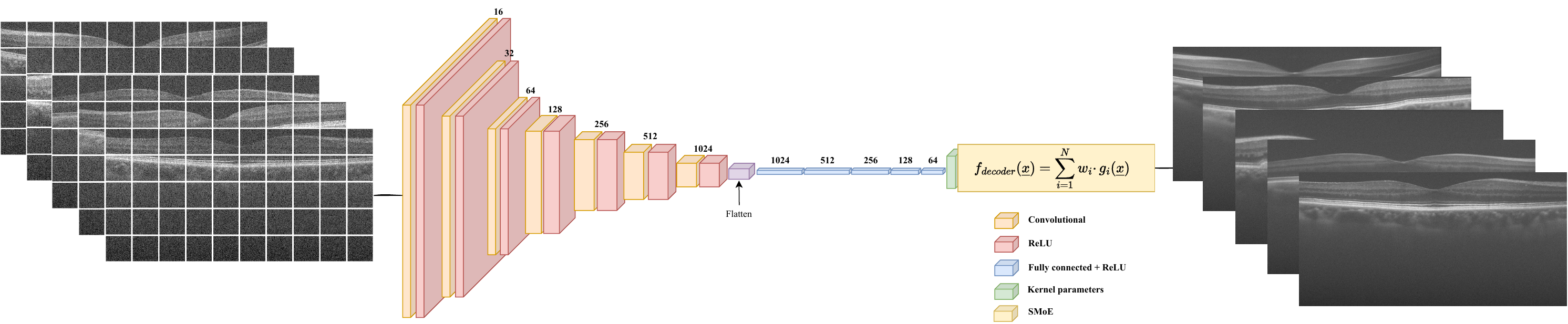}
   \end{tabular}
   \end{center}
   \caption[SMoE-GN] 
{ \label{fig:SMoE-GN} Encoder and Decoder (Gating) Network. }
\end{figure}

SMoE Gating Networks (SMoE-GN) which as depicted as Fig \ref{fig:SMoE-GN} might be expressed mathematically as follows:
\begin{equation}
\label{eq:f_encoder}
f_{encoder}\left(\underline{x}\right)=f_{FC}(\mathrm{Flatten}(f_{conv}^n(\ldots f_{conv}^2(f_{conv}^1(\underline{x})))))
\end{equation}

Here $f_{conv}^{\left(l\right)}\left(\cdot\right)$ represents the $l\mathrm{-th}$ convolutional layer, including the convolution operation, the addition of the bias, and the application of the nonlinearity (ReLU).

\begin{equation}
\label{eq:f_conv}
    f_{conv}^{\left(l\right)}\left(\mathbf{H}\right)=\mathrm{ReLU}({\mathrm{Conv}}^{\left(l\right)}\left(\mathbf{H}\right)+b^{(l)})
\end{equation}
$\mathbf{H}$ is the input of the $l\mathrm{-th}$ layer for $l=1,\ \mathbf{H}=\ \underline{x}$.  ${\mathrm{Conv}}^{\left(l\right)}$ is the convolution operation and $b^{\left(l\right)}$ is the bias term in the $l\mathrm{-th}$ layer.The $\mathrm{Flatten(\cdot)}$ operation reshapes the output of the final convolutional layer into a one-dimensional vector for each data sample, preparing it for the fully connected layers $f_{FC}(\cdot)$. These layers then serve as the bottleneck that combines the SMoE model parameters.
\begin{equation}
\label{eq:f_fc}
f_{FC}\left(\mathbf{H}\right)=\ \mathrm{ReLU}(\mathbf{W}^{\left(n\right)}\mathbf{H}+b^{\left(n\right)})
\end{equation}
\begin{gather}
    f_{decoder}\left(\underline{x}\right) = \sum_{i=1}^{N} w_i \cdot g_i\left(\underline{x}\right) \notag \\
    g_i\left(\underline{x}\right) = \frac{\pi_i \cdot k_i\left(\underline{x}\right)}{\sum_{i=1}^{N} \pi_i \cdot k_i\left(\underline{x}\right)}\ \ \ \text{with}\ \ \sum_{i=1}^{N} \pi_i = 1 \notag \\
    k_i(\underline{x}) = \exp\left(-\frac{1}{2}\left({\underline{x}}_i - {\underline{\mu}}_i\right)^T {{\underline{\Sigma}}_i}^{-1} \left({\underline{x}}_i - {\underline{\mu}}_i\right)\right)
     \label{eq:f_decoder}
\end{gather}

In Equation~\ref{eq:f_decoder}, the mixture of experts in the decoder is depicted, where $\pi_i$ represents the prior probability of the $i^{th}$ component, ensuring that the weights sum up to 1. The term ${\underline{\mu}_i}$ denotes the center of each expert's focus in a two-dimensional space, $\underline{\mu}_i=\left(\mu_{i,x}, \mu_{i,y}\right)$. The mixture weights $w_i$ can be interpreted as the impact that the $i^{th}$ expert is responsible for modeling the image at point ${\underline{x}}_i$. Thus, in this mixture model, $w_i$ signifies the importance of each expert, while $g_i(\underline{x})$ acts as a soft-gating mechanism.
The covariance matrix ${\underline{\Sigma}}_i \in \mathbb{R}^{2 \times 2}$, which appears in the third line of Equation~\ref{eq:f_decoder}, plays a crucial role. It acts as the 'steering' matrix, describing the bandwidth $\delta^2_{i,j}$ of the kernel for each dimension ($s_{i,j}, (i \neq j)$). This matrix will determine the 'spread' of the influence of each expert over the 2D space. The expression $\left({\underline{x}}_i - {\underline{\mu}}_i\right)^T {{\underline{\Sigma}}_i}^{-1} \left({\underline{x}}_i - {\underline{\mu}}_i\right)$ in the kernel function $k_i\left(\underline{x}\right)$ is the Mahalanobis distance, a measure of the distance between a point and a distribution, providing a measure of how 'far' a given point ${\underline{x}}_i$ is from the mean ${\underline{\mu}}_i$ under the specified covariance. The covariance matrix can be described as follows:
\begin{equation}
\label{eq:matrix}
{\underline{\Sigma}}_i=\left[\begin{matrix}\delta_{x_{1,1}}^2&s_{1,2}\\s_{2,1}&\delta_{x_{2,2}}^2\\\end{matrix}\right]
\end{equation}

Furthermore, Equation~\ref{eq:f_decoder} characterizes the SMoE model's gating network (GN), processing parameters estimated by the encoder as $f_{decoder} = f (\underline{x},\theta)$, with $\theta \leftarrow f_{encoder}(\underline{\widetilde{x}})$. The operation of the gating network is mathematically formulated as follows:
\begin{equation}
f_{ae}\left({\underline{\widetilde{x}}}_n\right)=\ \sum_{n=1}^{N}{f_{decoder}^{(n)}\left(f_{encoder}^{(n)}\left({\underline{\widetilde{x}}}_n\right)\right)}
\end{equation}
This relationship is essential for systematically modeling speckle noise in observed noisy image patches, represented as $\mathcal{D}=\left\{\underline{\widetilde{x}} : \underline{x} \times (1+\epsilon)\right\} \in \mathbb{P}$,each being a 3D group extracted using the BM algorithm (Sec.~\ref{sec:bm-alg}). To obtain the denoised signal $\hat{y}$, we apply a Multi-Model Fusion (MMF) strategy, as introduced in our previous work \cite{RN67}. This involves summing the outputs of $\mathbf{N}$ individual models, each denoted by $f_{ae}^n\left({\underline{\widetilde{x}}}_n\right)$.

\subsection{Training SMoE Gating Network}
As the objective function for the SMoE Gating Network, we utilized a composite loss\cite{RN102} comprising Mean Squared Error (MSE) and the Structural Similarity Index (SSIM), as depicted in Equation \ref{eq:loss}, differing from the approach in recent work \cite{RN68}. However, in our previous study\cite{RN103}, we employed SSIM as the loss function in GD for global optimization in image compression task.

\begin{equation}
\label{eq:loss}
\mathcal{L}=\ \mathcal{L}_{MSE}\cdot\lambda_{MSE}+\mathcal{L}_{SSIM}\cdot\lambda_{SSIM}
\end{equation}

In image restoration tasks, it is crucial to assess both exact reconstructions, as measured by the MSE, and the preservation of important features, as gauged by SSIM. By integrating these metrics, a more nuanced and robust assessment is achievable, guiding an improved optimization quality control process\cite{RN42}. 

Also, to be able to efficiently train the SMoE-GN depicted in Fig \ref{fig:SMoE-GN}. We used several techniques like Mixed Precision which operations are performed in both single (FP32) and half (FP16) precision. Gradient clipping is a technique to prevent exploding gradients in neural networks. And Learning Rate Scheduling is used to dynamically adjust based on composite loss. For our training and validation dataset, we utilized Google's Open Image Dataset.\cite{RN85}The RGB images were grayscaled and normalized. Training details are listed in the table as follows.

\begin{table}[ht]
\caption{Training Details for SMoE-GN} 
\label{tab:SMoE-GN}
\centering
\begin{tabular}{|l|c|}
\hline
Hyperparameter                      & Value                      \\ \hline
Train Patch (Block) Size            & \( 8 \times 8 \)           \\
Model Parameters of SMoE-GN       & \( 7.41 \times 10^7 \)     \\
Training Data Size                  & \( 2.048 \times 10^8 \)    \\
Validation Data Size                & \( 4.096 \times 10^7 \)    \\
Start/End Epochs                    & 50                         \\
Batch Size                          & 16                         \\
Learning Rate                       & \( 1 \times 10^{-3} \)     \\
Optimizer                           & Adam                       \\
Learning Rate Scheduler             & ReduceLROnPlateau          \\
Criterion                           & MSE+SSIM                   \\ \hline
\end{tabular}
\end{table}

\begin{table}[ht]
\caption{SMoE AE Encoder Comparisons} 
\label{tab:AE-Encoder-Comp}
\centering
\begin{tabular}{|l c l c c|}
\hline
Dataset & Size & Method & PSNR & SSIM \\ \hline
BM3D Dataset & \(512\times512\) & BM-SMoE-AE-v2 & 32.40 & 0.96 \\
BM3D Dataset & \(512\times512\) & BM-SMoE-AE-v1 & 31.68 & 0.94 \\
\hline
\end{tabular}
\end{table}

Tab.~\ref{tab:AE-Encoder-Comp} demonstrates the efficiency of the proposed training strategy with BM-SMoE-AE model. The encoder utilizing composite loss outperformed the previous model\cite{RN68}, achieving an average improvement of approximately $0.7$ dB in PSNR and $0.02$ in SSIM.

\section{RESULTS}
\subsection{Perceptual Quality Metrics for Reconstruction}
Before discussing the outcomes of the OCT denoising study, it is important to outline the metrics employed to assess the perceptual quality of the images. These metrics are categorized into full reference (FR) and no-reference (NR) perceptual quality indicators:
\begin{itemize}
    \item FR metrics include PSNR (Peak Signal-to-Noise Ratio), SSIM, LPIPS (Learned Perceptual Image Patch Similarity),\cite{RN3} and GMSD (Gradient Magnitude Similarity Deviation).\cite{RN4} These metrics facilitate a quantitative evaluation by comparing the denoised images to the original, undistorted ones.
    \item NR metrics consist of BRISQUE (Blind/Referenceless Image Spatial Quality Evaluator)\cite{RN82} and the CLIP (Contrastive Language-Image Pre-training) model\cite{RN80}. These metrics are designed to assess the quality of images without reference to the original, focusing on both the perceptual appearance and the intrinsic content of the images.
\end{itemize}
\subsection{Dataset}

The dataset employed in our experiment is publicly available\cite{RN88} and comprises images from the eyes of 28 individuals. This cohort encompasses a diverse selection of subjects, some of whom have non-neovascular age-related macular degeneration (AMD) and others who do not. Bioptigen, Inc. (Durham, NC, USA) provided the 840-nm SDOCT imaging system used to capture these images, that is noted for superior axial resolution. For every subject, a pair of detailed SDOCT scans were performed. The initial scan systematically mapped a square volume centered on the retinal fovea, incorporating 1000 A-scans per B-scan. 

We also applied our method to our in-house OCT dataset captured by a MHz Fourier Domain Mode Locked (FDML) laser system. This system is commercially available from OptoRes GmbH and has the following specifications: a 1.6 MHz A-line sweeping frequency for the FDML laser, a central wavelength of 1309 nm, and a bandwidth of 100 nm. The axial resolution of the system was determined to be 17 $\mu$m (in air) with the lateral resolution at the surface being 40 $\mu$m. Three volume sets of OCT images (B-scans) of kiwi fruit were acquired, with each B-scan composed of 1024 A-scans. During the OCT image reconstruction, the acquired interference signal was zero-padded, resulting in 1024 points in depth. The final resolution of each B-scan was 1024 by 1024 points. The ground truth OCT image was obtained by averaging 32 B-scans.

\subsection{Experimental Results and Analysis}

\begin{table}[ht!]
\caption{Quantitively denoising comparisons against the SotA methods for SDOCT Dataset\cite{RN88}} 
\label{tab:Result1}
\centering
\begin{tabular}{|m{1.5cm} m{2.0cm} m{1cm} m{1cm} m{1cm} m{1cm} m{1.3cm} m{1cm} m{1.4cm} m{1.4cm}|}
\hline
Size & Method & PSNR & SSIM & LPIPS & GSMD & BRISQUE & CLIP & Encoding Time(s) & Decoding Time(s) \\ \hline
450x900 & BM-SMoE–AE & \cellcolor{green!25}30.83 & \cellcolor{green!25}0.84 & 0.56 & 0.07 & 11.86 & 0.19 & 2.34 & 0.88 \\
450x900 & BM3D & 29.12 & 0.77 & 0.52 & 0.10 & 25.01 & 0.39 & N/A & N/A \\
450x900 & PBML-SMoE-AE & 28.30 & 0.67 & 0.64 & 0.15 & 28.70 & 0.46 & 45.92 & 1.83 \\
450x900 & DnCNN & 26.08 & 0.60 & 0.55 & 0.16 & 23.34 & 0.68 & N/A & N/A \\
450x900 & FFDNet & 24.47 & 0.54 & 0.54 & 0.17 & 48.78 & 0.72 & N/A & N/A \\
450x900 & Noisy & 18.31 & 0.31 & 0.55 & 0.24 & 0.00 & 0.00 & N/A & N/A \\ \hline
\end{tabular}
\end{table}

\begin{figure}[H]
   \begin{center}
   \begin{tabular}{c} 
   \includegraphics[height=6.4cm]{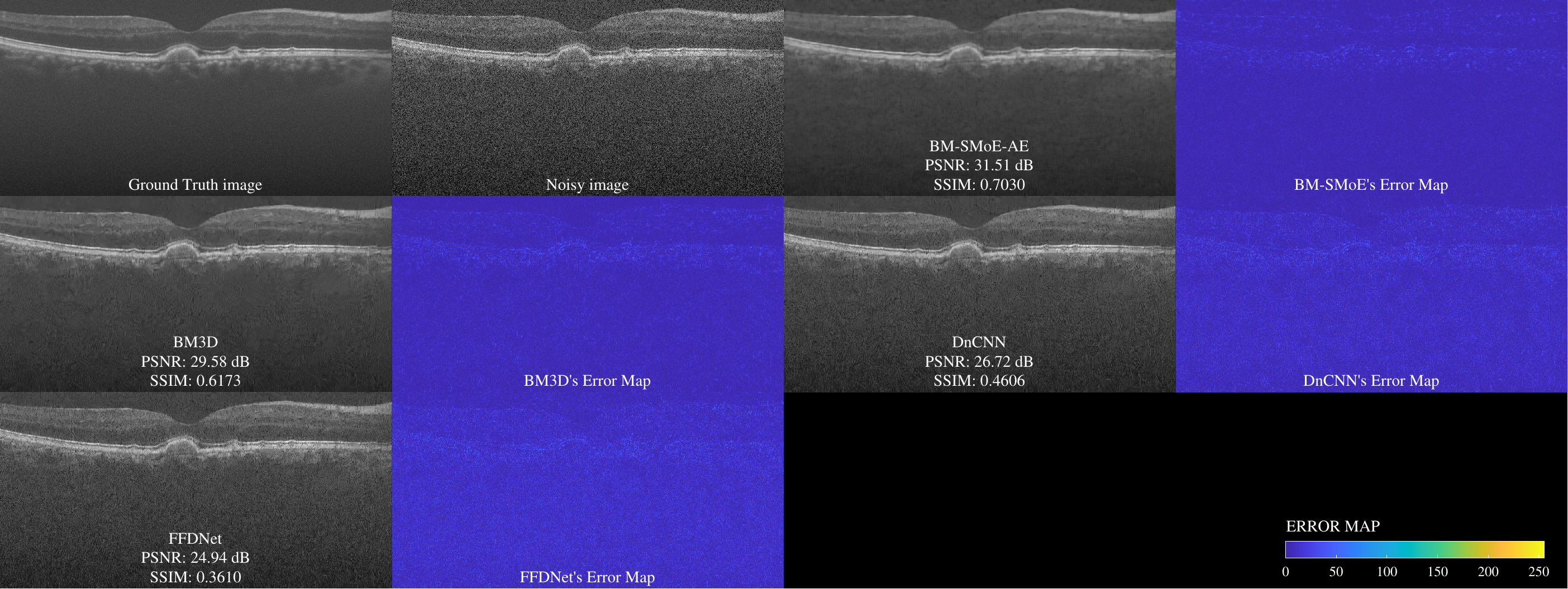}
   \end{tabular}
   \end{center}
   \caption[SBSDI-Result] 
{ \label{fig:Results-Vis1} Visual denoising comparisons against the SotA methods for SDOCT Dataset\cite{RN88}. }
\end{figure}
\begin{figure}[H]
   \begin{center}
   \begin{tabular}{c} 
   \includegraphics[height=8cm]{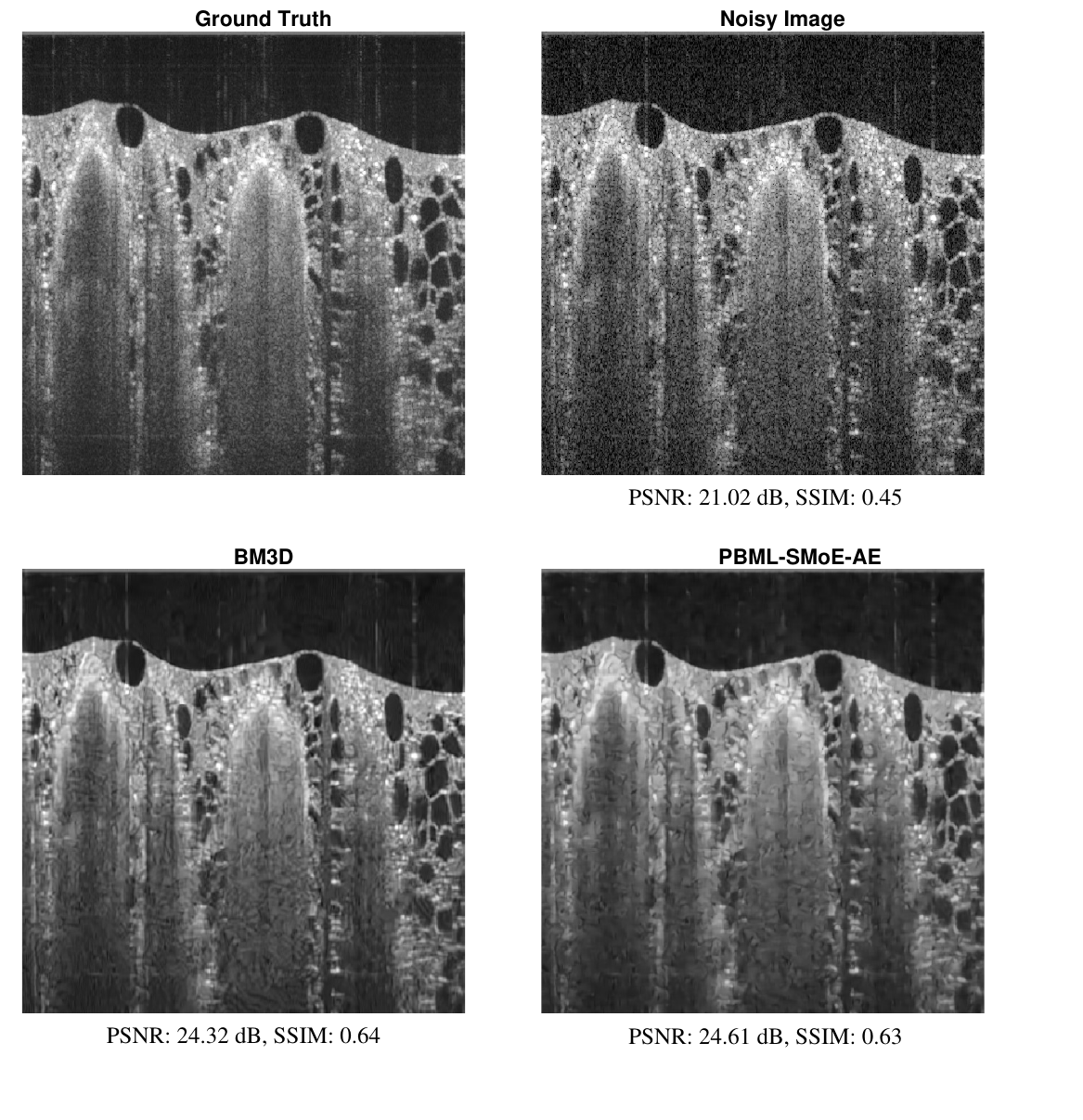}
   \end{tabular}
   \end{center}
   \caption[IOMT-Result] 
{ \label{fig:Results-Vis2} In house dataset, denoising performance comparison between PBML-SMoE and BM3D. }
\end{figure}

In the conducted experiments, as presented in Tab.~\ref{tab:Result1}, the BM-SMoE- AE algorithm demonstrated enhanced performance relative to both the BM3D and other state of the art methods. This was evidenced by its superior results in key metrics: it achieved an improvement in the PSNR of approximately 1.71 dB and there is an increment in the SSIM by 0.07 with respect to  BM3D. In Fig \ref{fig:Results-Vis1}, it is observed that BM-SMoE–AE surpasses BM3D with a margin of 1.93 dB in PSNR and 0.09 in SSIM. In Fig. \ref{fig:Results-Vis2}, we present a comparison between PBML-SMoE and BM3D using our in-house dataset. The results indicate that PBML-SMoE outperforms BM3D, achieving a higher PSNR by 0.29 dB and a marginally lower SSIM by 0.01. In addition to improvements in perceptual quality, the BM-SMoE model exhibits superior edge reconstruction capabilities compared to other methods. This is evidenced by our observations in Figures \ref{fig:Results-Vis2} and \ref{fig:Results-Vis1}.

The NR metrics we implemented, namely 'BRISQUE' and 'CLIP' (where lower scores indicate better quality), not only successfully identified high-quality reconstructions in alignment with other FR metrics but also showed potential in training neural networks in scenarios lacking ground truth data. This finding is particularly relevant in fields like OCT and medical imaging, where such metrics could be invaluable for evaluating image quality when a ground truth image is unavailable.

\section{CONCLUSION}
In this research paper, we extend the SMoE model's application to medical imaging, notably in OCT imaging for speckle noise reduction. This progression represents a notable advancement beyond conventional imaging techniques. We enhanced the model's efficiency by integrating neural network architecture with new training and an adaptive multi-model inference strategy. Nevertheless, the presence of millions of parameters in our approach renders it impractical and computationally burdensome. Recent research indicates that achieving state-of-the-art performance is feasible with a comparatively modest number of parameters. Furthermore, dealing with correlated image denoising, such as speckle, which can act as both noise and information in the data, remains a challenging task. This is particularly true when the ground truth images are not available, and the noise model is unknown. In our upcoming research, we will tackle these obstacles by exploring additional datasets. Our goal is to demonstrate the broad applicability of our algorithm while ensuring its resilience against correlated noise, all while keeping it lightweight and efficient.

\acknowledgments 
This project has received funding from the European Union’s Horizon 2020 research and innovation programme under the Marie Sklodowska-Curie grant agreement No 956770. This work is supported by COST Action CA21155 HISTRATE and Bulgarian National Science Fund under project КП-06-KOCT/19. V.M. and E.S. thank European Regional Development Fund within the Operational Programme “Science and Education for Smart Growth 2014–2020” under the Project CoE “National center of Mechatronics and Clean Technologies” BG05M2OP001-1.001-0008.

\bibliography{report} 
\bibliographystyle{spiebib} 

\end{document}